\documentstyle[12pt,epsfig]{article}
\def\figcap{\section*{Figure Captions\markboth
        {FIGURECAPTIONS}{FIGURECAPTIONS}}\list
        {Figure \arabic{enumi}:\hfill}{\settowidth\labelwidth{Figure
999:}
        \leftmargin\labelwidth
        \advance\leftmargin\labelsep\usecounter{enumi}}}
 \relax
\def\ap#1#2#3{Ann.\ Phys.\ (NY) #1 (19#3) #2}

\def\np#1#2#3{Nucl.\ Phys.\ B#1 (19#3) #2}
\def\pl#1#2#3{Phys.\ Lett.\ #1B (19#3) #2}
\def\pr#1#2#3{Phys.\ Rev.\ D #1 (19#3) #2}
\def\prb#1#2#3{Phys.\ Rev.\ B #1 (19#3) #2}
\def\prep#1#2#3{Phys.\ Rep.\ #1 (19#3) #2}

\def\rmp#1#2#3{Rev.\ Mod.\ Phys.\ #1 (19#3) #2}

\def\cmp#1#2#3{Comm.\ Math.\ Phys.\ #1 (19#3) #2}
\def\cmp#1#2#3{Comm.\ Math.\ Phys.\ #1 (19#3) #2}

\newcounter{hran}
\def\bmini{\setcounter{hran}{\value{equation}}
\refstepcounter{hran} \setcounter{equation}{0}
\renewcommand{\theequation}{\thehran\alph{equation}}
              \begin{eqnarray}  }
\def\bminia{\setcounter{hran}{\value{equation}}
\refstepcounter{hran} \setcounter{equation}{1}
\renewcommand{\theequation}{\thehran\alph{equation}}
              \begin{eqnarray}  }
\def\bminiG#1{
          \setcounter{hran}{\value{equation}}
          \refstepcounter{hran}
          \setcounter{equation}{-1}
          \renewcommand{\theequation}{\thehran\alph{equation}}
          \refstepcounter{equation}
    \label{#1}
          \begin{eqnarray}          }
\def\emini{\end{eqnarray}\setcounter{equation}{\value{hran}}
\renewcommand{\theequation}{\arabic{equation}}}
\def\half{\mbox{\small $\frac{1}{2}$}}

\def\frac#1#2{ {{#1} \over {#2} }}

\def\tr{\mbox{tr}}
\def\fun#1#2{\lower3.6pt\vbox{\baselineskip0pt\lineskip.9pt
  \ialign{$\mathsurround=0pt#1\hfil##\hfil$\crcr#2\crcr\sim\crcr}}}
\def\ie{\hbox{\it i.e.}{ }}

\def\ds#1{\ooalign{$\hfil/\hfil$\crcr$#1$}}
\def\re#1{(\ref{#1})}
\def\beq{\begin{equation}}
\def\eeq{\end{equation}}
\def\beeq{\begin{eqnarray}}
\def\eeeq{\end{eqnarray}}
\def\beeqn{\begin{eqnarray*}}
\def\eeeqn{\end{eqnarray*}}
\def\bc{\bar c}
\def\bp{\bar p}

\def\bpsi{\bar \psi}
\def\r{\rho}
\def\s{\sigma}
\def\S{\Sigma}
\def\G{\Gamma}
\def\eps{\epsilon}

\def\L{\Lambda}
\def\l{\lambda}
\def\bl{\bar \lambda}
\def\g{ \gamma}
\def\de{\delta}

\def\z{\zeta}
\def\d4#1{\frac {d^4 {#1} }{(2\pi)^4}}

\def\UV{$\L_0\to\infty\;$}
\def\IR{$\L=0\;$}
\def\bit{\begin{itemize}}
\def\eit{\end{itemize}}
\def\ben{\begin{enumerate}}
\def\een{\end{enumerate}}

\def\nome#1{{\label{#1}}}
\def\p{\partial}

\begin{document}
\begin{titlepage}
\renewcommand{\thefootnote}{\fnsymbol{footnote}}
\begin{flushright}
     UPRF 94-392 \\
     February 1994
\end{flushright}
\par \vskip 10mm
\begin{center}
{\Large \bf
Axial anomalies in gauge theory by \\
exact renormalization group  method
\footnote{Research supported in part by MURST, Italy}}
\end{center}
\par \vskip 2mm
\begin{center}
        {\bf M.\ Bonini, M.\ D'Attanasio} \\
        Dipartimento di Fisica, Universit\`a di Parma and\\
        INFN, Gruppo Collegato di Parma, Italy\\
        and\\
        {\bf G.\ Marchesini}\\
        Dipartimento di Fisica, Universit\`a di Milano and\\
        INFN, Sezione di Milano, Italy
\end{center}
\par \vskip 2mm
\begin{center} {\large \bf Abstract} \end{center}
\begin{quote}
The global chiral symmetry of a $SU(2)$ gauge theory is
studied in the framework of renormalization group (RG).
The theory is defined by the RG flow equations in the infrared
cutoff $\L$ and the boundary conditions for the relevant couplings.
The physical theory is obtained at $\L=0$.
In our approach the symmetry is implemented by choosing
the boundary conditions for the relevant couplings not at the
ultraviolet point $\L=\L_0\to\infty$ but at the physical value $\L=0$.
As an illustration, we compute the triangle axial anomalies.
\end{quote}
\end{titlepage}

\noindent
In a gauge theory the introduction of a momentum cutoff
seems to conflict with the local symmetry and gives rise to the fine
tuning problem \cite{Becchi,romani}.
A solution of this problem for a Yang-Mills (YM) theory has been proposed
in \cite{YM}, by using an approach based on the renormalization group (RG)
method \cite{W}-\cite{BDM}. In this formulation one works directly
in four space-time dimensions and thus its extension to
theories with chiral symmetry is straightforward.
This is an advantages with respect to dimensional regularization
\cite{dr}, in which one has to define
$\g_5$ in complex space-time dimensions.

In this paper we study a $SU(2)$ non-Abelian gauge theory with two fermions.
Neglecting the fermion masses this theory has the global
$U(1)_V \times U(1)_A \times SU(2)_V \times SU(2)_A$ symmetry with four
classically conserved currents. This symmetry gives rise to Ward
identities for the vertices with current insertions.
By using the RG method we study, in perturbation theory,
these Ward identities and compute the corresponding anomalies
\cite{Anomaly} for
the three point vertices, which require only a one loop calculations.

According to the RG procedure we follow, the needed subtractions are
generated by imposing the usual physical conditions on all
the relevant parameters of the vertex considered. The question now is
how to choose the physical condition for a vertex with current
insertions. For each vertex one has a given number of parameters and
Ward identities. One then selects the relevant parameters in such a
way to satisfy some of the Ward identities. If there are more Ward
identities than parameters, then the remaining ones are
predicted and typically anomalous.

Before recalling the RG method and the corresponding loop expansion,
we describe the theory, its symmetries and Ward identities.
We consider the model in which the BRS action in the Feynman gauge is
\beeq\nonumber
S_{BRS}=\int d^4x\biggl\{
-\frac 1 4 \left( F_{\mu\nu}  \cdot F_{\mu\nu}\right)
-\half  \left(\p_\mu W_\mu \right)^2
+ w_\mu \cdot D_\mu c -\half v \cdot c \wedge c \\
+\sum_{f=1}^2\left(\bpsi_f(i\ds{D}-m_f)\psi_f
-\bl_f \,c\cdot t\,\psi_f-\bpsi_f \,c\cdot t\,\l_f
\right)\biggr\}\,, \nonumber
\eeeq
where
$$
F^a_{\mu\nu}(x)=\p_\mu W^a_\nu -\p_\nu W^a_\mu
+ g \left( W_\mu \wedge W_\nu\right)^a \,,
\;\;\;\;\;\;
\left(W_\mu \wedge W_\nu\right)^a=
\epsilon^{abc} W^b_\mu W^c_\nu \,,
$$
$$
F_{\mu\nu} \cdot F_{\mu\nu} =F_{\mu\nu} ^a F_{\mu\nu} ^a\,,
\;\;\;\;\;
D_\mu c =\p_\mu c + gW_\mu \wedge c\,,
\;\;\;\;\;
D_\mu \psi_f =(\p_\mu + gW_\mu\cdot t)\psi_f\,,
\;\;\;\;\;
w_\mu=\frac 1 g u_\mu+\p_\mu \bc\,,
$$
$f$ is the flavour index and $t^a$ are $SU(2)$ matrices in the fundamental
representation. This action is BRS invariant and
we have added the sources $u_\mu^a$, $v^a$, $\l_f$ and $\bl_f$ for the BRS
transformations \cite{BRS} of $W_\mu^a$, $c^a$, $\bpsi_f$ and $\psi_f$
respectively.
For $m_f\to 0$ this action has the additional global
$U(1)_V \times U(1)_A \times SU(2)_V \times SU(2)_A$ symmetry in the flavour
space, with the four classically conserved currents
$$
j^A_\mu=\left\{
\bpsi\g_\mu\psi\,,\;\;
\bpsi\g_\mu\g_5\psi\,,\;\;
\bpsi\g_\mu\tau^\alpha\psi\,,\;\;
\bpsi\g_\mu\g_5\tau^\alpha\psi\right\}\,.
$$
We denoted by $\tau^\alpha$ the three $SU(2)$ matrices acting on the
two dimensional flavour
space, in order to avoid confusion with the colour matrices $t^a$.
The index $A$ indicates the flavour symmetry group, namely
$A\;=\;\bigl\{1,2,3\alpha,4\alpha\bigr\}$.
At the quantum level the properties of these currents are studied by
introducing the sources $\z^A_\mu$
\beq\nome{Sbrs'}
S^\z_{BRS}=S_{BRS}+\int d^4x\z^A_\mu j^A_\mu \,.
\eeq
{}From $S^\z_{BRS}$ we can construct the effective action
$\G[\Phi]$, where $\Phi(x)$ denotes all the fields and sources, namely
$\Phi=(W_\mu^a,\psi,\bpsi,c^a,\bc^a,u_\mu^a,v^a,\l,\bl,\z^A)$.
This functional satisfies the Slavnov-Taylor
identities
\beq\nome{st}
\int d^4x \left\{
\frac 1 g
\frac{\de\G'}{\de W_\mu^a}
\frac{\de\G'}{\de w_\mu^a}+
\frac{\de\G'}{\de c^a}
\frac{\de\G'}{\de v^a}-
\frac{\de\G'}{\de \bpsi_f}
\frac{\de\G'}{\de \l_f}+
\frac{\de\G'}{\de \psi_f}
\frac{\de\G'}{\de \bl_f}
\right\}=0\,,
\eeq
where $\G'=\G+\half\int d^4x(\p_\mu W^a_\mu)^2$.
On the other side, the Ward identities associated to the chiral symmetry
transformations are in general anomalous. Namely, defining the functional
\beq\nome{wi}
{\cal W}^A=\int d^4x \left\{
-i\p_\mu\frac{\de\G}{\de \z_\mu^A}+\frac{\de\G}{\de \psi}
T^A\psi+\bpsi{\bar T}^A \frac{\de\G}{\de \bpsi}+f^{ABC}\z^B_\mu
\frac{\de\G}{\de \z^C_\mu}
\right\} \,,
\eeq
we have
$$
{\cal W}^A={\cal A}^A+{\cal O}(m_f)\,,
$$
where the ${\cal O}(m_f)$ corrections are present only for $A=2$ or $4\alpha$
(\ie for $U(1)_A$ or $SU(2)_A$) and for  $A=3\alpha$ (\ie for
$SU(2)_V$) if $m_f$ is not independent of $f$.
In \re{wi} we have set
$T^A=\bigl\{1,\g_5,\tau^\alpha,\g_5\tau^\alpha\bigr\}$,
${\bar T}^A=\bigl\{ 1,-\g_5,\tau^\alpha,-\g_5\tau^\alpha \bigr\}$ and
$f^{ABC}$ is
different from zero only if $(A,B,C)\;=\;(3\alpha,3\beta,3\g)$,
$(3\alpha,4\beta,4\g)$, $(4\alpha,3\beta,4\g)$ or
$(4\alpha,4\beta,3\g)$.
In these cases $f^{ABC}=-i\eps^{\alpha\beta\g}$.
As well known the anomaly ${\cal A}^A$ can be set to zero for $A=1$ and $3$
(\ie for $U(1)_V$ and
$SU(2)_V$). For $A=U(1)_A$ or $SU(2)_A$ there are some anomalous
contributions.

The physical parameters (masses, wave function renormalization and coupling
constants) are given by fixing at some normalization point
the vertex functions with non-negative dimension and possibly their
derivatives. A similar thing has to be done for the vertices involving the
currents $j_\mu^A$.
Also for these one has to identify the relevant parameters (with
non-negative dimension) and the irrelevant vertices (with negative
dimension). The relevant parameters are present in
the following contribution to the effective action
\beeq\nome{Grel}
&&\G^\z[W,\bpsi,\psi,\z]=\int d^4x \biggl\{
\z^A_\mu \bigl[ \bpsi O^{(\z\bpsi\psi)A}_\mu
\psi+O_{\mu\nu\r}^{Aab}W^a_\nu W^b_\r+ \nonumber
O_{\mu\nu\r\s}^{Aabc}W^a_\nu W^b_\r W^c_\s \bigr] \\
\;\;\;\;\;\;\;\;&& +\z^A_\mu \z^B_\nu \bigl[
\half O_{\mu\nu}^{AB}+O_{\mu\nu\r\s}^{ABab}W^a_\r W^b_\s\bigr]
+\z^A_\mu \z^B_\nu \z^C_\r O_{\mu\nu\r}^{ABC}
+\z^A_\mu \z^B_\nu \z^C_\r \z^D_\s O_{\mu\nu\r\s}^{ABCD}\biggr\}
\,,
\eeeq
which contains only vertices $O$ with non-negative dimension.
We define  the relevant parameters only for the
three point vertices, since they give the basic anomalies.
The decomposition for the other vertices in \re{Grel} can be done in an
analogous way.
We will work in the momentum space.
We denote the fermion momenta by the letter $k$, the gluon
momenta by $q$ and the current momenta by $p$.
For simplicity we fix the subtraction point at vanishing momenta. This
is possible for $m_f\ne 0$. In general one can use a symmetric
subtraction point ($3SP$), defined by
$\bar p_i \bar p_j=\frac{\mu^2}{2}(3\delta_{ij}-1)$, with some
complications in the definition of the vertices.

\noindent
1) The vertex $O^{Aab}_{\mu\nu\r}$ is present only for $A=2$
(\ie $U(1)_A$), due to charge conjugation invariance, and we have
$$
O_{\mu\nu\r}^{Aab}(p,q,q')=\de^{A2}\,\de^{ab}\left\{
\eps_{\mu\nu\r\s}\,(q-q')_\s[\r_5+\S_5(p,q,q')]
+\S_{\mu\nu\r}^5(p,q,q')\right\}\,,
$$
$$
\S_5(0,0,0)=0\,.
$$
{}From this condition $\S_5(p,q,q')$ is irrelevant (after factorizing a
momentum factor, this vertex has negative dimension).
In $\S_{\mu\nu\r}^5(p,q,q')$ the
Lorentz indices are carried by the momentum, then this vertex is
irrelevant.
The ST identities \re{st} imply for this vertex the transversality of
the gluons
\beq\nome{tran}
q_\nu O_{\mu\nu\r}^{Aab}(p,q,q')=q'_\r O_{\mu\nu\r}^{Aab}(p,q,q')=0\,.
\eeq
{}From \re{wi} one has to consider also the vertex
$$
{\cal W}_{\nu\r}^{Aab}(p,q,q')=p_\mu O_{\mu\nu\r}^{Aab}(p,q,q')\,,
$$
which, as known, contains an anomaly.

\noindent
2) For the $O^{ABC}_{\mu\nu\r}$ vertex we have to distinguish five
types of contributions, corresponding to the different values of the
indices $ABC$ which give a nonvanishing result.

(a) For $(A,B,C)=(1,3\beta,4\g)$ we have ($c^{ABC}=\de^{\beta\g}$)
$$
O_{\mu\nu\r}^{ABC}(p,p',p'')=c^{ABC}\,\biggl\{
\eps_{\mu\nu\r\s}\,p'_\s\;[\r_6^{ABC}+\S_6^{ABC}(p,p',p'')]
+\eps_{\mu\nu\r\s}\,p''_\s\;[\r_7^{ABC}+\S_7^{ABC}(p,p',p'')]
$$
\beq\nome{ro6}
+\S_{\mu\nu\r}^{ABC}(p,p',p'')\biggr\}\,,
\;\;\;\;\;\;\;\;\;\;\;\;\;\;\;\;
\S_6(0,0,0)=\S_7(0,0,0)=0\,.
\eeq
For $(A,B,C)=(2,1,1)$, $\;(2,3\beta,3\g)$ and $(2,4\beta,4\g)$, we have
the same decomposition with $\r_6=-\r_7$ and $\S_6=-\S_7$, due to the
symmetry under the exchange of $B$ and $C$, and with
$c^{ABC}=1$, $\de^{\beta\g}$ and $\de^{\beta\g}$, respectively.

(b) For $(A,B,C)=(2,2,2)$ the coefficient proportional to one momentum
vanishes for symmetry reasons and we have an irrelevant vertex
$$
O_{\mu\nu\r}^{ABC}(p,p',p'')=\S_{\mu\nu\r}^{ABC}(p,p',p'')\,.
$$

(c) For $(A,B,C)=(3\alpha,3\beta,3\g)$ we have
\beeqn
O_{\mu\nu\r}^{ABC}(p,p',p'')=\eps^{\alpha\beta\g}\,\biggl\{[
g_{\mu\nu}(p'-p)_\r+
g_{\nu\r}(p''-p')_\mu+
g_{\mu\r}(p-p'')_\nu]\,
[\r_8+\S_8(p,p',p'')]
\\
+\S_{\mu\nu\r}^{ABC}(p,p',p'')\biggr\}\,,
\;\;\;\;\;\;\;\;\;\;\;\;\;\;\;\;\;\;\;\;\;\;\;\;\;
\S_8(0,0,0)=0\,.
\eeeqn

(d) For $(A,B,C)=(3\alpha,4\beta,4\g)$ we have
\beeqn
&& O_{\mu\nu\r}^{ABC}(p,p',p'')=\eps^{\alpha\beta\g}\,\biggl\{
g_{\nu\r}(p''-p')_\mu
[\r_{9}+\S_{9}(p,p',p'')]+(g_{\mu\nu} p_\r-g_{\mu\r} p_\nu)\\
&&\;\; \times
[\r_{10}+\S_{10}(p,p',p'')]
+(g_{\mu\nu} p'_\r-g_{\mu\r} p''_\nu)
[\r_{11}+\S_{11}(p,p',p'')]
+\S_{\mu\nu\r}^{ABC}(p,p',p'')\biggr\}
\,,
\eeeqn
$$
\S_{9}(0,0,0)=\S_{10}(0,0,0)=\S_{11}(0,0,0)=0\,.
$$
{}From \re{wi} the vertex of ${\cal W}^A$ which contains
$O_{\mu\nu\r}^{ABC}$ is
$$
{\cal W}_{\nu\r}^{ABC}(p,p',p'')=
p_\mu O_{\mu\nu\r}^{ABC}(p,p',p'')
+f^{ABX}\,O_{\nu\r}^{XC}(p'')
+f^{ACX}\,O_{\nu\r}^{XB}(p')
\,.
$$
As known, the axial identities are anomalous in the cases (a) and (b).

The situation for the three point vertices is summarized in tab.~1, where
we report for each vertex the number of relevant parameters and the relations
given by ST or Ward identities.

We have now to fix the values of these $10$ relevant parameters. First of all
we must require the ST identities \re{tran}, which involve only the vertex
$O_{\mu\nu\r}^{Aab}$. This implies $\r_5=0$.
Then, following the usual procedure, we require as many Ward identities
${\cal W}^A=0$ as possible.
However, as shown in the table, one has for the three point vertices more
relations than relevant parameters.
Therefore some of the relations ${\cal W}^A=0$ can not be satisfied.
One usually prefers to satisfy the vector Ward identities.
In this case one obtains that all the $10$ couplings $\r$ are zero.
This choice is given in tab.~1, which shows that the Ward
identities which should be evaluated and that may contain anomalies are
\beeqn
&&{\cal W}_{\nu\r}^{ABC}(p,p',p'')\,,
\;\;\;\;\;\;\;\;
\;\;\;\;\;\;\;\;
ABC=211,233,244,413,222\,, \\
&&{\cal W}_{\nu\r}^{Aab}(p,k,k')\,,
\;\;\;\;\;\;\;\;
\;\;\;\;\;\;\;\;\;\;
A=2\,.
\eeeqn
In the next section we show how this procedure of extracting the
relevant couplings naturally emerges in the framework of exact RG.

\begin{table}
$$
\begin{array}{cc||c||c|c|c|c|c||c}
\multicolumn{2}{c||}{\mbox{vertices}}&\mbox{parameters}&
\multicolumn{5}{c||}{\mbox{relations}}&\mbox{predictions} \\
&&& ST & U(1)_V & U(1)_A & SU(2)_V & SU(2)_A & \\ \hline \hline
%\multicolumn{2}{c||}{O_\mu^A}	&4& &1&1&1&1&- \\ \hline
%\multicolumn{2}{c||}{O_L^A}   &4+4& &1&1&1&1&- \\ \hline
%\multicolumn{2}{c||}{O_T^A}	&4&
%\multicolumn{5}{c||}{\mbox{no tree level kinetic term}}&-
%\\ \hline
O_{\mu\nu\r}^{Aab} 	&A=2		&1&1& &1& & &U(1)_A \\ \hline
			&ABC=211	&1& &1&1& & &U(1)_A  \\ \cline{2-9}
			&ABC=233	&1& & &1&1& &U(1)_A  \\ \cline{2-9}
			&ABC=244	&1& & &1& &1&U(1)_A  \\ \cline{2-9}
O_{\mu\nu\r}^{ABC}	&ABC=413	&2& &1& &1&1&SU(2)_A  \\ \cline{2-9}
			&ABC=222	&-& & &1& & &U(1)_A  \\ \cline{2-9}
			&ABC=333	&1& & & &1& &- \\ \cline{2-9}
			&ABC=344	&3& & & &1&1&- \\ \hline
\end{array}
$$
\caption{\small{Parameters and Ward identities for the three-point
vertices}}
\end{table}

\vspace{3mm}\noindent{\bf 1. RG formulation}

\noindent
The exact RG formulation for this theory can be obtained by
generalizing the method used for the $SU(2)$ Yang-Mills case in
ref.~\cite{YM}. For completeness we recall here the procedure and
the relevant new steps.
\newline
{\it i) Cutoff effective action.}
The ``cutoff effective action'' $\G[\Phi;\L,\L_0]$,
is obtained by putting an infrared (IR) cutoff $\L$ and an ultraviolet
(UV) cutoff $\L_0$ for all propagators in the vertices of
the physical effective action $\G[\Phi]$. Namely one sets to zero
each propagator if its frequency is lower than $\L$ or larger than $\L_0$.
Then $\G[\Phi;\L=0,\L_0\to\infty]$ is just the physical effective action
and, at this point, the ST identities and the Ward identities,
anomalous or not, should be satisfied.
\begin{figure}
\begin{center}
\mbox{\epsfig{file=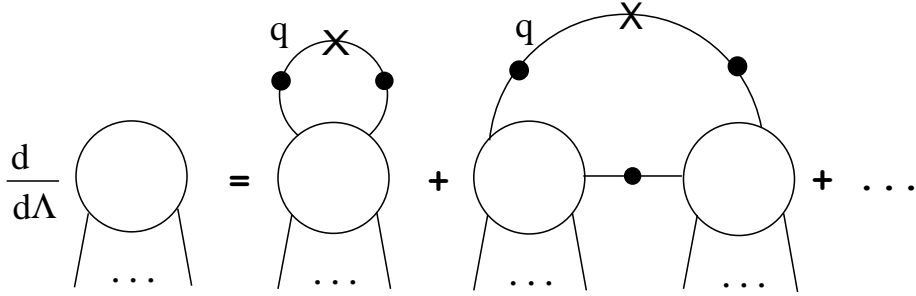,height=4cm}}
\end{center}
\caption{\small{
The circles represent vertex functions with the IR cutoff $\L$.
Internal lines involve only the fields $W_\mu,c,\bar c,\psi,\bpsi$,
while the external lines could involve also the
sources $u_\mu$, $v$, $\l$, $\bl$ and $\z^A$.
The crosses represent the derivative with respect to $\L$ of the internal
propagators. Integration over $q$ in the loop is understood.
}}
\end{figure}
\newline
{\it ii) Evolution equation.}
The RG flow in the IR cutoff $\L$ is more easily written
for the interacting part $\Pi[\Phi;\L,\L_0]$, \ie the cutoff effective
action $\G[\Phi;\L,\L_0]$ minus the free inverse cutoff propagator
contributions. The evolution equation has the form
\beq\nome{eveq}
\L\partial_\L \Pi[\Phi;\L,\L_0]=I[\Phi;\L,\L_0]\,,
\eeq
where the functional $I[\Phi;\L,\L_0]$ is given (non linearly) in terms of
$\G[\Phi;\L,\L_0]$.
This equation is obtained by observing that $\L$ enters only as a
cutoff in all internal propagators and is pictorially given in
fig.~1. The expression of $I[\Phi;\L,\L_0]$ can be obtained form
the one given in ref.~\cite{YM} for the pure YM case by adding
the fermion field contribution. As shown in fig.~1 it is constructed
in terms of vertices of $\G[\Phi;\L,\L_0]$. This equation is suitable
for the loop expansion, due to the fact that the $q$-integration in
the r.h.s. adds a loop.
\newline
{\it iii) Relevant couplings.}
The evolution equation \re{eveq} allows one to compute the cutoff effective
action once the boundary conditions are given.
As one expects, the boundary conditions for the various vertices
depend on dimensional counting. Therefore
one must distinguish irrelevant vertices, which have negative mass
dimension, and relevant couplings, \ie couplings with non-negative dimension.
In order to identify these relevant parameters
(masses, wave function constants and couplings),
one has to consider all the vertices of $\G[\Phi;\L,\L_0]$ with
non-negative dimension.
There are $10$ of such vertices involving only the fields and sources
$W_\mu$, $\bpsi$, $\psi$, $c$, $w_\mu$, $v$, $\bl$ and $\l$.
The extraction of the relevant couplings for the Yang-Mills sector is done
in ref.~\cite{YM}, and for the fermion sector in the appendix.
For the part involving the sources $\z_\mu^A$, one has to consider the $7$
vertices $O$ in \re{Grel}, which now depend on $\L$.
{}From them one extracts the relevant parameters $\r_i(\L)$ and the
irrelevant vertices $\S_i(p,\cdots;\L)$, as previously done.
After this one divides the effective action in two parts,
$\Pi=\Pi_{rel}+\G_{irr}$, where $\Pi_{rel}$ is given by a
polynomial in $\Phi$ with these relevant couplings as coefficients.
The irrelevant part $\G_{irr}$ of the cutoff effective action contains
the vertices with negative dimension and the vertices $\S_i$.
\newline
{\it iv) Boundary conditions.}
The boundary conditions for the evolution equation \re{eveq} are fixed as
follows:
(a) Due to dimensional reason, the irrelevant part of the effective action
is fixed to vanish at $\L=\L_0\to \infty$, namely we impose the
boundary condition
\beq\nome{bc1}
\G_{irr}[\Phi;\L=\L_0,\L_0]=0 \,;
\eeq
(b) The boundary conditions on the relevant part of the effective action
are given at \IR and fix the physical parameters.
As discussed in ref.~\cite{YM}, they are
\beq\nome{bc2}
\Pi_{rel}[\Phi;0,\L_0]=S^{\z int}_{BRS}
+\int d^4x\, \biggl\{
\frac{\r_{4A}(0)}{8} \left[2(W_\mu\cdot W_\nu)^2 +(W_\mu\cdot W_\mu)^2
\right]+\frac{\r_{vcc}(0)}{2} v \cdot c \wedge c
\biggr\}\,,
\eeq
where $S^{\z int}_{BRS}$ is obtained from \re{Sbrs'} after having subtracted
the free inverse propagators.
The two additional couplings $\r_{4A}(0)$ and $\r_{vcc}(0)$ are due to
a four gluon quantum interaction with a group structure not present in
$S^\z_{BRS}$ and quantum corrections to the BRS variation of $c$.
Notice that one does not have a corresponding correction for the
contribution of the BRS variation of $W_\mu$ since the effective
action depends on the
combination $w_\mu=\frac 1 g u_\mu+\p_\mu c$.
The presence of these additional terms with couplings $\r_{4A}(0)$ and
$\r_{vcc}(0)$ is directly associated to the fact that the ST identities
are nonlinear and mix relevant and irrelevant part of the action. Since
$S^\z_{BRS}$ satisfies by itself ST identities, it follows that the two
couplings must be fixed in terms of irrelevant vertices evaluated at
appropriated normalization points. This is explained in detail in
\cite{YM}.
As shown in the appendix, in the quark sector there is no such mixing.
\newline
{\it v) Loop expansion.}
By using the boundary conditions \re{bc1} and \re{bc2},
the evolution equation \re{eveq} can be written as an integral equation
\beq\nome{inteq}
\Pi[\Phi;\L,\L_0]=\Pi_{rel}[\Phi;0,\L_0]+\int_0^\L \frac{d\l}{\l}
I[\Phi;\l,\L_0]+\int_0^{\L_0} \frac{d\l}{\l}
\left\{ I[\Phi;\l,\L_0]-I_{rel}[\Phi;\l,\L_0] \right\} \,,
\eeq
where the relevant part $I_{rel}$ of the functional $I$, defined
in \re{eveq}, is identified in the same way as $\Pi_{rel}$.
The iterative solution of this equation gives the usual loop expansion.
In the last term of \re{inteq} the functional $I_{rel}$
gives the subtractions needed to have a finite \UV limit.
The first loop calculation for the pure YM sector are reported in
\cite{YM}, together with the analysis of some ST identities
in the physical limit ($\L=0$ and \UV).
In the next section we perform the one loop calculations for the
current sector and show how the vector Ward identities are satisfied in the
physical limit, while the axial ones are anomalous.

\vspace{3mm}\noindent{\bf 2. One loop calculations}

\noindent
In order to show that this formulation is suitable for dealing with
chiral symmetries (global in this case) we compute the anomalies.
The calculation is simple and similar to the one performed by Taylor
expansion of Feynman diagram integrands.

The one loop calculations are deduced from the iterative solution of
\re{inteq}. Since we are interested in the physical vertices, we set $\L=0$.
At zero loop one has $\G[\Phi]=S^\z_{BRS}$.
As shown in \cite{BDM}, the one loop contribution
$\G^{(1)}(p_1\cdots p_n)$ of a general $n$-point vertex, is obtained
as follows.
First one computes $\G'(p_1\cdots p_n,\L_0)$, which
is the usual one loop Feynman graph, in which all internal propagators have an
UV cutoff $\L_0$. If the vertex $\G(p_1\cdots p_n)$ has negative dimension,
the loop integration is convergent and one has $\G^{(1)}(p_1\cdots p_n)=
\G'(p_1\cdots p_n,\L_0\to\infty)$. If $\G(p_1\cdots p_n)$ has non-negative
dimension, one must impose the physical conditions on the relevant
parameters. This generates the necessary subtractions and the limit \UV can
be taken.

We now show that the vector Ward identities are satisfied if we fix
all the $\r(0)=0$ in the current sector. Moreover, one finds the
correct values of anomalies for the axial Ward identities.
We limit our analysis to the three point functions.
For simplicity we set $m_f=m$ independent of $f$.

We consider first the case which gives the axial
anomaly, namely $O_{\mu\nu\r}^{ABC}$ with $(A,B,C)=(2,1,1)$.
The one loop graph is
\beeq
O'_{\mu\nu\rho}(p,p',p'')=i \int_\ell Tr
\frac{K_{\L_0}(\ell)}{\ds \ell -m} \g_\mu \g_5
\frac{K_{\L_0}(\ell-p'-p'')}{\ds \ell -\ds p' -\ds p'' -m}
\g_\rho \frac{K_{\L_0}(\ell-p')}{\ds \ell -\ds p' -m} \g_\nu \nonumber \\
+\;\;(p',\nu)\to(p'',\rho)\,,\nonumber
\eeeq
with $K_{\L_0}(\ell)$ rapidly vanishing for $\ell^2 \geq \L_0^2$ and
$\int_\ell=\int\frac{d^4\ell}{(2\pi)^4}$.
Imposing the condition $\r_6(0)=0$ we have
$$
O^{211}_{\mu\nu\r}(p,p',p'')=
O'_{\mu\nu\r}(p,p',p'')
-(p'-p'')_\s\,
\frac{\partial}{\partial \bp'_\s}
O'_{\mu\nu\r}(\bp,\bp',\bp'')|_{\bp'=\bp''=0}\,.
$$
We now proceed in evaluating the Ward identities for this vertex. The
calculation follows the usual steps, in particular we must
take into account the surface terms coming from differences of cutoff
functions. We find
$$
p'_\nu O'_{\mu\nu\rho}(p,p',p'') = -p''_\nu O'_{\mu\rho\nu}(p,p',p'')
=\frac{1}{6\pi^2}
\eps_{\mu\alpha\rho\beta}p'_\alpha p''_\beta+{\cal O}(\frac{1}{\L_0^2})\,,
$$
\beq\nome{W5'}
p_\mu O'_{\mu\nu\rho}(p,p',p'') = \frac{1}{6\pi^2}
\eps_{\alpha\nu\rho\beta}p'_\alpha p''_\beta+{\cal O}(m)
+{\cal O}(\frac{1}{\L_0^2})\,.
\eeq
{}From this we obtain
$$
\frac{\partial}{\partial \bp'_\s}
O'_{\mu\nu\r}(\bp,\bp',\bp'')|_{\bp'=\bp''=0}=
-\frac{1}{6\pi^2} \eps_{\mu\nu\r\s}
+{\cal O}(\frac{1}{\L_0^2})\,.
$$
Thus we have that the vector Ward identities are satisfied in the limit \UV,
while, as well known, at one loop the axial Ward identity develops an
anomalous term
$$
p'_\nu\,O^{211}_{\mu\nu\rho}(p,p',p'')=
p''_\nu\,O^{211}_{\mu\rho\nu}(p,p',p'')=0\,,
\;\;\;\;\;\;
p_\mu\,O^{211}_{\mu\nu\rho}(p,p',p'')=\frac{1}{2\pi^2}
\eps_{\alpha\nu\rho\beta} p'_\alpha p''_\beta +{\cal O}(m)\,.
$$
In the limit $m\to 0$ we can write the form of the physical vertex
$O^{211}_{\mu\nu\rho}$ in terms of the dimensionless function $\S_6^{211}$,
defined in \re{ro6}
\beeq
&&O^{211}_{\mu\nu\rho}(p,p',p'')=\frac{1}{2\pi^2}
\eps_{\alpha\nu\rho\beta} \frac{p'_\alpha p''_\beta p_\mu}{p^2}+
\biggl[ \nonumber
\eps_{\mu\nu\rho\alpha} (p'-p'')_\alpha+
\eps_{\mu\nu\alpha\beta} \frac{p'_\alpha p''_\beta p_\r}{p\cdot p''}+
\\
&&\eps_{\mu\alpha\r\beta} \frac{p'_\alpha p''_\beta p_\nu}{p\cdot p'}-
2\eps_{\alpha\nu\r\beta} \frac{p'_\alpha p''_\beta p_\mu}{p^2}
\biggr]\; \S_6^{211}(p,p',p'') \,.\nonumber
\eeeq
Notice that, as discussed by \cite{BS}, this amplitude is not a
pure pole in $p^2$.

The case of $O^{Aab}_{\mu\nu\rho}$ is analogous to the previous one. In fact
by choosing $\r_5(0)=0$, we have
$$
O^{Aab}_{\mu\nu\rho}(p,p',p'')=-g^2\de^{A2}\, \tr (t^a t^b)
O^{211}_{\mu\nu\rho}(p,p',p'')\,.
$$
Thus the axial Ward identity is anomalous.

Also in the two cases $(A,B,C)=(2,3\alpha,3\beta)$ and
$(A,B,C)=(2,4\alpha,4\beta)$, setting $\r^{ABC}_6(0)=0$, we
fulfil the vector Ward identities but we find an anomalous axial contribution.
We have
\beq\nome{a2}
O^{ABC}_{\mu\nu\rho}(p,p',p'')=\tr (\tau^\alpha \tau^\beta)
O^{211}_{\mu\nu\rho}(p,p',p'')\,.
\eeq

When $(A,B,C)=(1,3\alpha,4\beta)$, apart for a factor
$\tr (\tau^\alpha \tau^\beta)$, the unsubtracted vertex is given by
$O'_{\mu\nu\rho}(p,p',p'')$. The one loop vertex is then
($\r_6(0)=\r_7(0)=0$)
$$
O^{13\alpha 4\beta}_{\mu\nu\r}(p,p',p'')= \tr (\tau^\alpha \tau^\beta)
\biggl\{ O'_{\mu\nu\r}(p,p',p'')
-\bigl( p'_\s \frac{\partial}{\partial \bp'_\s}
+p''_\s \frac{\partial}{\partial \bp''_\s} \bigr)
O'_{\mu\nu\r}(\bp,\bp',\bp'')|_{\bp'=\bp''=0} \biggr\} \,.
$$
{}From \re{W5'} one finds
$$
\frac{\partial}{\partial \bp'_\s}
O'_{\mu\nu\r}(\bp,\bp',\bp'')|_{\bp'=\bp''=0}=-
\frac{\partial}{\partial \bp''_\s}
O'_{\mu\nu\r}(\bp,\bp',\bp'')|_{\bp'=\bp''=0}=
-\frac{1}{6\pi^2} \eps_{\mu\nu\r\s}
+{\cal O}(\frac{1}{\L_0^2})\,.
$$
Thus for this vertex we have the same result as in \re{a2}.

If $(A,B,C)=(2,2,2)$, since no subtraction is needed we have that
$O^{222}_{\mu\nu\r}=O'_{\mu\nu\r}$ and find from \re{W5'} that all the
axial identities are anomalous.

In the case $(A,B,C)=(3\alpha,4\beta,4\g)$ the Ward identities
${\cal W}^A=0$ are
\beeq
&& p_\mu
O^{3\alpha 4\beta 4\g}_{\mu\nu\r}(p,p',p'')=
i \eps^{\alpha\beta\g'} O^{4\g' 4\g}_{\nu\r}(p'')+
i \eps^{\alpha\g\beta'} O^{4\beta' 4\beta}_{\nu\r}(p')\,, \nonumber\\
&& p'_\nu
O^{3\alpha 4\beta 4\g}_{\mu\nu\r}(p,p',p'')=
i \eps^{\beta\g\alpha'} O^{3\alpha' 3\alpha}_{\mu\r}(p)+
i \eps^{\beta\alpha\g'} O^{4\g' 4\g}_{\mu\r}(p'') \,.\nonumber
\eeeq
These imply $\r_{9}=\r_{10}=\r_{11}=0$ at $\L=0$.
The Ward identities for this vertex are analogous to the three gluon vertex
ST identity and we will not verify it.

In the case $(A,B,C)=(3\alpha,3\beta,3\g)$ the Ward identity
${\cal W}^A=0$ is
$$
p_\mu
O^{3\alpha 3\beta 3\g}_{\mu\nu\r}(p,p',p'')=
i \eps^{\alpha\g\beta'} O^{3\beta' 3\beta}_{\nu\r}(p')+
i \eps^{\alpha\beta\g'} O^{3\g' 3\g}_{\nu\r}(p'')\,.
$$
This implies $\r_8(0)=0$. At one loop this vertex and the corresponding
Ward identity are equal to the previous case
$(A,B,C)=(3\alpha,4\beta,4\g)$.

In this paper we analyzed a RG formulation of a non-Abelian gauge theory
with global chiral symmetry. In dimensional regularization \cite {dr}
such a study is technically difficult due to the noninvariant
counterterms induced by the definition of $\g_5$ in complex dimensions
\cite{chi}. This problem is known as the fine tuning problem.

When studying the RG evolution equation for a local gauge theory, one
meets the same problem, since the momentum cutoff is a
symmetry-breaking regulator \cite{Becchi,romani}.
In a previous paper \cite{YM} we showed
how this difficulty can be avoided by imposing the physical boundary
conditions on the relevant part of the cutoff effective action.
In this way the bare couplings are automatically fine-tuned.

Since in the RG framework one works directly in four dimensions,
it should be straightforward to extend our procedure to local chiral symmetry.
This is confirmed by the analysis of the global chiral symmetry we
have made in this paper.
This symmetry is implemented when one gives the boundary conditions
for the relevant part of the cutoff effective action in the current sector.
In sect.~2 we performed the one loop calculations, analyzed in this
formulation
the axial Ward identities for the three current vertices and
re-obtained the anomalies given by their well-known values.
The one loop calculation of the chiral Ward identity involving the vertex
$O_\mu^{(\z\bpsi\psi)A}$, performed in the appendix, also shows that we indeed
avoid the complicated fine tuning of the counterterms (see for instance
ref.~\cite{dr}).

This formulation provides a systematic cutoff
procedure which generates higher order corrections.
This is done by solving iteratively the integral equation \re{inteq}.
This equation is in principle nonperturbative, although only iterative
solution has been considered here. This is another
advantage respect to dimensional regularization, which is perturbative in
nature and, for chiral theories, difficult to extend to higher orders.

We have benefited greatly from discussions with A. Bassetto, C.
Becchi, R. Soldati, M. Testa and M. Tonin.

\vspace{3mm}\noindent{\bf Appendix}

In this appendix we analyze the fermion sector. First of all we
decompose the vertices involving the quarks into relevant and
irrelevant parts.

\noindent
1) The contribution to $\G[\Phi;\L,\L_0]$ from the quark propagator,
$\bpsi_f(-k)\G^{(\bpsi\psi)}(k;\L)\psi_f(k)$,
can be written as
$$
\G^{(\bpsi\psi)}(k;\L)=(\ds k -m_f)K^{-1}_{\L\L_0}(k)+\s_{m_\psi}(\L)+
\s_{\psi}(\L)(\ds k -m_f)+\S^{(\bpsi\psi)}(k;\L)\,,
$$
$$
\S^{(\bpsi\psi)}(k;\L)=0\;\;\;\;\;\mbox{at}\;\;\;\ds k =m_f\,,
\;\;\;\;\;\;\;\;\;\;\;\;\frac{\p}{\p k_\mu}
\S^{(\bpsi\psi)}(k;\L)=0\;\;\;\;\;\mbox{at}\;\;\;k^2=\mu^2\,.
$$
2) For the quark-gluon contribution,
$\bpsi(k)\G^a_\mu(k,k',q;\L)\psi(k')W^a_\mu(q)$, we define
$$
\G^a_\mu(k,k',q;\L)=\g_\mu t^a \s_{g}(\L)+\S^{(\bpsi\psi W)a}_\mu(k,k',q;\L)\,,
\;\;\;\;\;\;\;\;\;\;\;\;
\S^{(\bpsi\psi W)a}_\mu(\bar k ,-\bar{k},0;\L)|_{\bar{k}^2=\mu^2}=0\,.
$$
3) For the contribution $\bl(p)\G^{(\bl\psi c)a}(p,k,q;\L)\psi(k)c^a(q)$,
involving the source $\bl$, we define
$$
\G^{(\bl\psi c)a}(p,k,q;\L)=t^a \s_{\bl}(\L)+\S^{(\bl\psi c)a}(p,k,q;\L)\,,
\;\;\;\;\;\;\;\;\;\;\;\;\S^{(\bl\psi c)a}(p,k,q;\L)|_{3SP}=0\,,
$$
and analogously for $\G^{(\bpsi\l c)a}$. All the $\s$'s are relevant
couplings while the $\S$'s are irrelevant vertices.
The contribution of the above vertices to the relevant part of the
cutoff effective action is then ($\s_i=\s_i(\L)$)
$$
\Pi^{F}_{rel}[\Phi;\L]=\int d^4x \; \biggl\{
\bpsi_f\left[\s_{m_\psi}+\s_{\psi}(i\ds \p -m_f)\right]\psi_f+
\s_{g} \bpsi_f \ds W \cdot t\, \psi_f
+ \s_{\bl} \left[ \bl_f c \cdot t\,\psi_f
+\bpsi_f c \cdot t\,\l_f  \right] \biggr\}\,.
$$
The boundary conditions for the five parameters $\s$ are fixed
according to eq.~\re{bc2}. The fact that the coupling $\s_{\bl}$ does
not receive corrections can be seen by considering
the ST identity for the quark-gluon vertex at \IR
$$
\G_\mu^a(k,k',q)\G^{(uc)}_\mu(q)+
\G^{(\bpsi\l c)a}(k,k',q)\G^{(\bpsi\psi)}(k')-
\G^{(\bl\psi c)a}(k,k',q)\G^{(\bpsi\psi)}(-k)
=0\,.
$$
By evaluating this identity at $3SP$, one finds that all the irrelevant
contributions vanish, and the coupling $\s_{\bl}(0)$ is fixed to be its tree
level value ($\s_{\bl}(0)=1$).

We now examine the only relevant vertex which couples fermions and
currents, namely the first vertex in \re{Grel}. We have
$$
O_\mu^{(\z\bpsi\psi)A}(p,k,k')=\left\{\g_\mu(1+\r_1^A)+\S_\mu^A(p,k,k')
\right\}T^A\,, \;\;\;\;\;\;\S_\mu^A(0,\bar k,-\bar k)|_{\bar k^2=\mu^2}=0
\,.
$$
{}From \re{wi}, the vertex of the functional ${\cal W}^A$ which contains
$O_\mu^{(\z\bpsi\psi)A}$ is
$$
{\cal W}^{(\z\bpsi\psi)A}(p,k,k')=p_\mu
O_\mu^{(\z\bpsi\psi)A}(p,k,k')-
\G^{(\bpsi\psi)}(-k)T^A+ \bar T^A \G^{(\bpsi\psi)}(k') \,.
$$
We now show how this identity ${\cal W}^A=0$ is satisfied in the
physical limit at one loop order. Thus we set \IR and, for simplicity,
$m_f=m$.

At zero loop we have $O_\mu^{(\z\bpsi\psi)A}=\g_\mu T^A$ and
$\G^{(\bpsi\psi)}(k)=\ds{k}-m$, thus the Ward identity
${\cal W}^{(\z\bpsi\psi)A}=0$ is trivially
satisfied in the limit $m\to 0$ at this order.

(a) One loop calculation of $\G^{(\bpsi\psi)}(k)=\ds{k}-m+\Pi(k)$.
The one loop graph is ($C_F=\frac 3 4$)
$$
\Pi'(k)=-i g^2 C_F \int_\ell \frac 1 {\ell^2}
\g_\rho \frac{1}{\ds k+\ds \ell -m} \g_\rho
K_{\L_0}(\ell)K_{\L_0}(k+\ell)\,.
$$
Imposing the physical renormalization conditions, we find
$$
\Pi(k)=\Pi'(k)-\Pi'(\bar k )-
(k_\mu-{\bar k }_\mu)\partial_{{\bar k'}_\mu}\Pi'({\bar k'})
\,,\;\;\;\;\;\;\ds{\bar k}=m\,,
\;\;\;\bar{k'}^2=\mu^2\,.
$$
Because of these subtractions, we can take the limit \UV. The one loop
fermion propagator is then
$$
\Pi(k)=-i g^2 C_F \int_\ell
\frac 1 {\ell^2}
\left\{
\g_\rho \left( \frac{1}{\ds k+\ds \ell -m}-
\frac{1}{\ds{\bar k}+\ds \ell -m} \right) \g_\rho
+(k_\mu-{\bar k }_\mu)\, \chi_\mu(\bar{k'},\bar{k'},\ell) \right\}\,,
$$
where
$$
\chi_\mu(p,p',\ell)\equiv \g_\rho
\frac 1 {\ds \ell +\ds p - m} \g_\mu \frac 1 {\ds \ell +\ds p' -m}\g_\rho\,.
$$

(b) One loop calculation of $O_\mu^{(\z\bpsi\psi)A}$. The one loop graph is
$$
O'^A_\mu(p,k,k')=ig^2 C_F \int_\ell\frac 1 {\ell^2} \, \chi^A_\mu(k',-k,\ell)
K_{\L_0}(\ell) K_{\L_0}(\ell+k') K_{\L_0}(\ell-k) \,,
$$
with
$$
\chi^A_\mu(p,p',\ell)\equiv \g_\rho
\frac 1 {\ds \ell +\ds p - m} \g_\mu T^A \frac 1 {\ds \ell +\ds p'
-m}\g_\rho\,.
$$
Imposing the condition $\r_1^A(0)=0$ we find, in the limit \UV,
the one loop contribution to $O_\mu^{(\z\bpsi\psi)A}$
\beeq
O_\mu^{(\z\bpsi\psi)A}(p,k,k')&=&O'^A_\mu(p,k,k')-O'^A_\mu(0,\bar k'
,-\bar k' )
\nonumber \\
&=& ig^2 C_F \int_\ell\frac 1 {\ell^2} \left[ \chi^A_\mu(k',-k,\ell)
- \chi^A_\mu(-\bar k',-\bar k',\ell)\right]\,.\nonumber
\eeeq
Using
$$
p_\mu\, \chi_\mu^A(k',-k,\ell)=\g_\rho\left(
\frac 1 {\ds \ell +\ds {k'} -m}{\bar T}^A -T^A \frac 1 {\ds \ell -\ds k -m}
\right)\g_\rho +{\cal O}(m)\,,
$$
where the ${\cal O}(m)$ term is present if $A=2,4\alpha$. Then
one automatically satisfies at one loop level the Ward identity
${\cal W}^{(\z\bpsi\psi)A}=0$ in the limit $m\to 0$.

\end{document}